\documentclass{iopconfser}

\usepackage{graphicx}
\usepackage{float}
\usepackage{amsmath}
\usepackage[backend=biber, style=numeric, doi=true, url=true, maxnames=1, maxbibnames=1, sorting=none]{biblatex}
\usepackage{booktabs}

\begin{document}

\title{Inverse Surrogate Model of a Soft X-Ray Spectrometer using Domain Adaptation}

\author{Enrico Ahlers$^{*,1,2}$, Peter Feuer-Forson$^{*,1}$, Gregor Hartmann$^{1}$, Rolf Mitzner$^{1}$, Peter Baumgärtel$^{1}$ and Jens Viefhaus$^{1}$}

\affil{$^1$Helmholtz-Zentrum für Materialien und Energie GmbH, Albert-Einstein-Straße 15, 12489 Berlin, Germany}
\vspace{20pt}
\affil{$^2$Humboldt-Universität zu Berlin, Unter den Linden 6, 10099 Berlin, Germany}
\vspace{20pt}
\affil{$^*$Both have equally contributed as first authors}
\email{peter.feuer-forson@helmholtz-berlin.de, enrico.philip.ahlers@hu-berlin.de}
\vspace{10pt}
\begin{abstract}
In this study, we present a method to create a robust inverse surrogate model for a soft X-ray spectrometer. During a beamtime at an electron storage ring, such as BESSY II, instrumentation and beamlines are required to be correctly aligned and calibrated for optimal experimental conditions. In order to automate these processes, machine learning methods can be developed and implemented, but in many cases these methods require the use of an inverse model which maps the output of the experiment, such as a detector image, to the parameters of the device. Due to limited experimental data, such models are often trained with simulated data, which creates the challenge of compensating for the inherent differences between simulation and experiment. In order to close this gap, we demonstrate the application of data augmentation and adversarial domain adaptation techniques, with which we can predict absolute coordinates for the automated alignment of our spectrometer. Bridging the simulation-experiment gap with minimal real-world data opens new avenues for automated experimentation using machine learning in scientific instrumentation.
\end{abstract}

\section{Introduction}
Beamline experiments conducted at electron storage rings and other large-scale research facilities are both costly and in high demand. Consequently, it is essential to minimise the time spent on setup, calibration, and instrumentation alignment during experiments. Current research efforts are focused on automating these processes to improve efficiency. Work has already been performed to automate the process of an alignment method for a spectrometer which utilises a reflection zone plate (RZP) as a diffractive element \cite{Feuer-Forson:ys5108}. In this paper, a method was demonstrated which applied a surrogate model, trained with simulated data, in order to align a single RZP on a planar substrate. The method achieves the required inversion by means of optimisation, establishing offsets between the simulation and the experiment by comparing outputs of a forward-direction neural network and the equivalent detector images taken by the spectrometer. Our aim is to work towards improving this method by directly predicting absolute positions via an inverse surrogate, which takes as input the detector image and predicts directly the parameters of the spectrometer. This method has the potential to significantly reduce the run-time of the alignment process by foregoing the need to utilise an optimisation algorithm.

Here we present an inverse surrogate model for a soft X-ray spectrometer utilising a multi-RZP optical element. The primary motivation is to enable its use during beamline setup and alignment by accurately interpreting experimental images. However, direct training on experimental data is infeasible due to limited availability; in this case, 1,620 experimental samples were manually acquired in one week of beamtime, whereas to train the network 100,000 data points were required. To address this, we train the model using simulated data and apply augmentation and domain adaptation techniques \cite{2022GloTP, Csurka2017, 8099799} to bridge the gap between synthetic and experimental domains. This approach ensures the model can handle noise, artefacts, signal variations, and distortions, assisting researchers in optimizing experimental conditions, such as achieving a small focus at the detector position.

\section{Background}
\subsection{Soft X-ray Spectrometer}
\begin{figure}[H]
    \centering
    \includegraphics[width=0.6\linewidth]{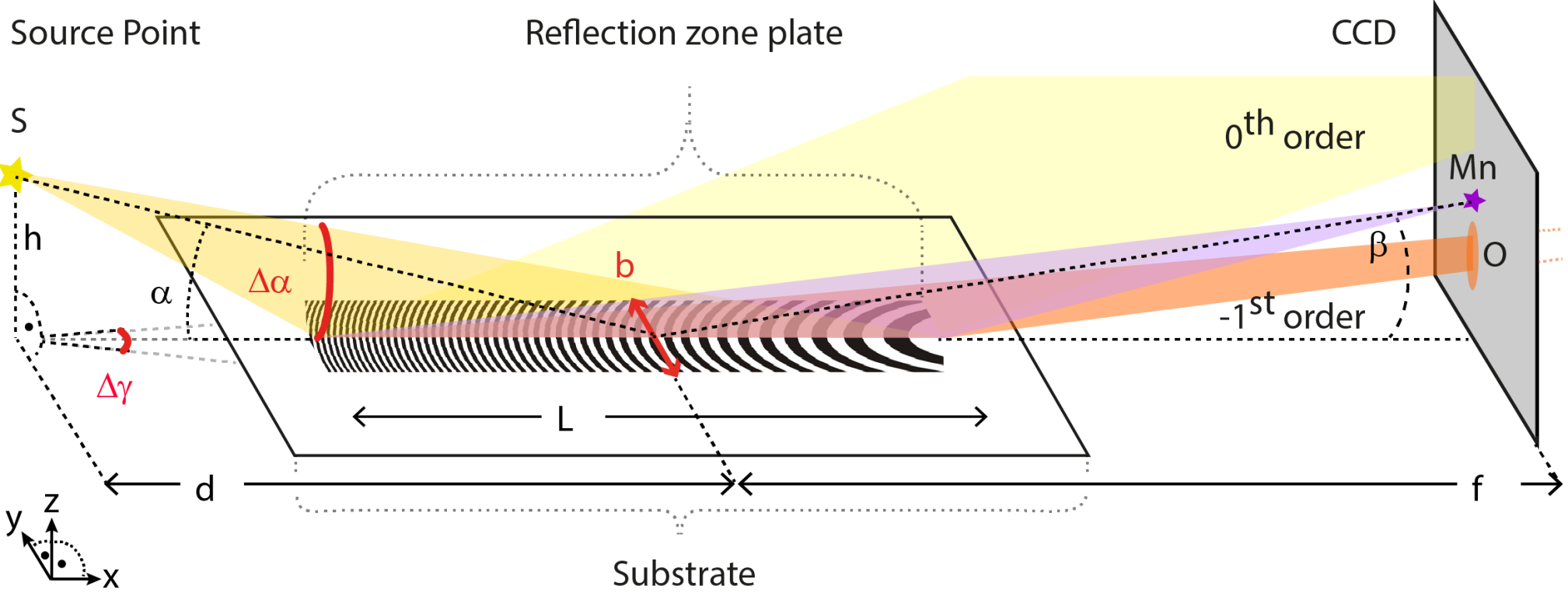}
    \caption{Principle of an RZP, \cite{kubin_manganese_2019}. In our spectrometer, the distance d is 90 mm, and the focal length f is 400 mm. The sample is a 4 mm-long wire composed of Cu\(86\), Mn\(12\), and Ni\(2\). The CCD image shows the diffraction and focusing of Mn, along with the separation achieved for oxygen and 0th-order light.}
\label{RZP}
\end{figure}

As part of our ongoing Röntgen-Ångström Cluster (RÅC) project, we are investigating various possible applications of machine learning methods, to assist the design and development of a new soft X-ray spectrometer for partial-fluorescence yield X-ray absorption spectroscopy at the 3d transition-metal L-edges. This work is important for many fields, for example, catalysis. The spectrometer consists of three chambers: the first houses the sample, the second contains the optical element—a planar substrate with an array of 100 RZPs—and the third holds the detector. A slit between the sample and the RZP minimises scattering. The task of the RZP is to detect faint signals, such as manganese fluorescence, even when the manganese-to-oxygen ratio is significantly skewed towards oxygen \cite{10.1063/1.4986627}. When performing an experiment, we connect our spectrometer to an open port at BESSY II and align the optics and camera relative to the sample. This alignment is necessary every time either the optics or the sample are changed. Figure \ref{RZP} illustrates the principle of the RZP and depicts the setup of our spectrometer.

\subsection{Simulation}
In order to train surrogate models of our spectrometer, it is necessary to generate synthetic data of the required quantity to train a deep neural network, and therefore efficient simulation software is required. At Helmholtz-Zentrum Berlin (HZB), we are currently developing a new software package for the simulation of beamlines and instrumentation called RAYX \cite{rayx}. This ray tracing software utilises modern GPU hardware in order to improve the run-time of simulations significantly and afford the possibility to create large datasets of millions of simulations for the production of surrogate models. These surrogate models provide approximate, yet ideally accurate results, and are ideal substitutes for computationally intensive problems.

\section{Method}
\subsection{Dataset Generation and Augmentation}
The dataset was created by running 100,000 simulations using RAYX with randomly chosen values for the \textit{x}- and \textit{y}-positions of the slit, with the \textit{x}-interval being \([-0.7, -0.5]\) and the \textit{y}-interval \([-1.4, -1.0]\). The \textit{y}-position of the image plane (detector) was randomised in the interval \([0.0, 2.0]\). These measurements are all in millimetres and correspond with the actual spectrometer. The number of rays were also varied per simulation between 500,000 and 1,000,000. These random settings are required to create a robust dataset which can account for variance within the setup of the spectrometer. Intervals for the \textit{x}-, \textit{y}-, and \textit{z}-positions of the RZPs and detector relative to the sample were chosen to match the experimental measurements. These intervals represent the maximum distances the motors can travel before no further signal reaches the detector (for \textit{x} and \textit{y}). These three parameters represent the output of the model. The values were randomly sampled from the following intervals, in millimetres: \textit{x}-axis: \([-3.8, 4.3]\), \textit{y}-axis: \([-3.8, 3.4]\), and \textit{z}-axis: \([83.0, 96.0]\).

To assist the training process, the dimensionality of the detector image was reduced by binning by a factor of 16, resulting in a 128 by 128 pixel monochrome image. The completed dataset contains input-output pairs, where Y consists of the \textit{x}-, \textit{y}- and \textit{z}-positions of the spectrometer and X is the resulting 128 by 128 pixel 2D histogram. A filtering process was undertaken to remove empty images, and therefore only samples with a minimum of 200 rays reaching the image plane are included. All features and the histogram are normalised to the range 0 to 1.

\begin{figure}[H]
    \centering
    \includegraphics[width=0.6\linewidth]{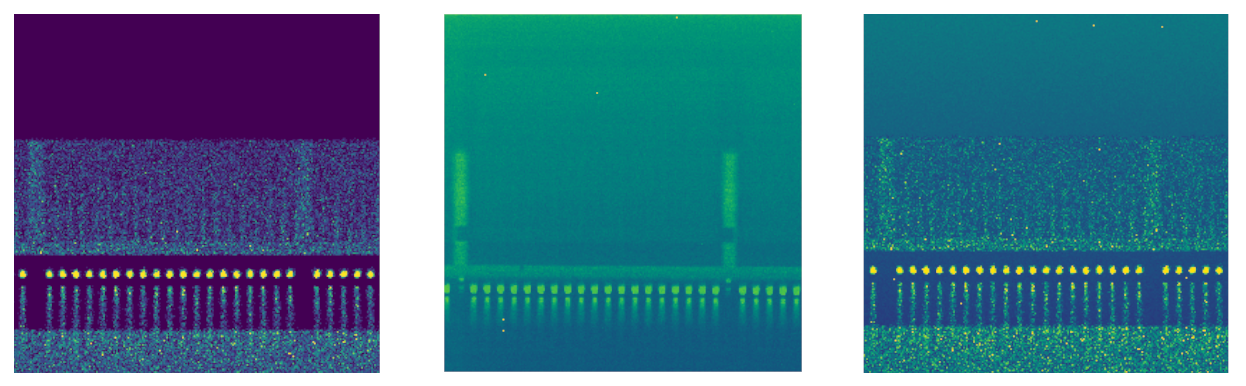}
    \caption{The left image shows the simulation, the middle image represents the experiment, and the right image depicts the simulation with augmentations applied.}
\label{augmentations}
\end{figure}

To address the variance between simulation and experiment, generate additional training samples, improve the network's ability to generalise to unseen data, and prevent overfitting, the following data augmentations were applied: random brightness, colour jitter, noise, rotation, hot pixels, Gaussian blur, and a data-specific noise gradient, which imitates the shift in intensity present in the experiment images. Figure \ref{augmentations} shows a simulated image directly from RAYX, an equivalent image captured by the spectrometer under experimental conditions, and an example of a simulated image with the data augmentations applied.

\subsection{Inverse Surrogate}
The goal of our work is to create a robust inverse surrogate model for a soft X-ray spectrometer with a RZP optical element, which uses 100 zone plates in an array. The challenge is creating a surrogate that inverts the simulation and takes as input experimental images. Given an image that was created during a real world experiment, the configuration of the \textit{x}-, \textit{y}- and \textit{z}-position of the spectrometer should be determined.

Neural networks excel at learning patterns and relationships within a specific dataset or domain (the range of data on which they are trained), but they often struggle to achieve accurate and reliable results on samples outside this domain. To assist the network in avoiding overfitting to the training set and generalise better to unseen data, data augmentation can be applied \cite{2022GloTP}. However, when the source and target datasets are highly divergent, augmentation alone may struggle to overcome the dissimilarity and a technique called domain adaptation can be used to align the domains and minimise the domain shift \cite{Csurka2017}. There are various methods to achieve this, one of which is called adversarial domain adaptation \cite{8099799}. This can be applied via an adversarial training methodology, where two neural networks compete against each other in a minimax game. In this game, a domain discriminator learns to distinguish between data from the source and target datasets. The feature extractor then works against the discriminator by learning domain invariant features and, in essence, tries to trick the discriminator. If the feature extractor achieves this then the latent representation it generates can be used by a regressor, to solve the given problem, which in our case is predicting parameters of our spectrometer.

The method we have implemented utilises two models which are trained simultaneously through a process known as adversarial training. These models are typically called the generator and the discriminator:

\begin{itemize}
    \item \textbf{Generator}: The generator's objective is to create data that closely mimics real data. In our case, it transforms input images to a latent representation.
    \item \textbf{Discriminator}: The discriminator's task is to determine whether a given data sample transformed by the generator originated from the experimental or simulation dataset, and outputs a probability score that reflects this.
\end{itemize}

During training, the generator and discriminator engage in this minimax game, whereby the generator tries to minimise the discriminator's ability to correctly classify its outputs, while the discriminator works to maximise its accuracy in distinguishing between simulated and experiment data.

\begin{figure}[H]
    \centering
    \includegraphics[width=0.9\linewidth]{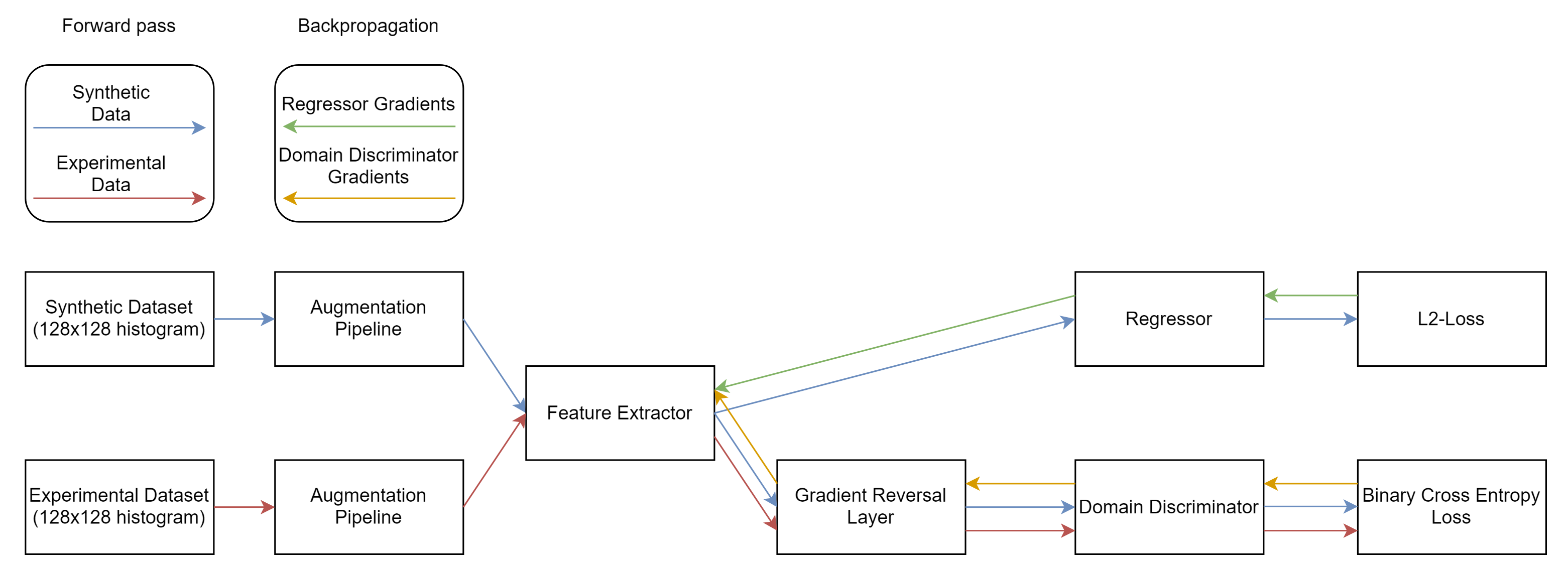}
    \caption{Structure of the adversarial domain adaptation network, showing the data paths for both synthetic and experimental data. The feature extractor is a convolutional network, the regressor and domain discriminator are fully-connected networks.}
\label{network}
\end{figure}

Firstly, the images are processed by the feature extractor, which is a convolutional network which acts as the generator. This module extracts latent features from the images. This representation is then flattened into a one-dimensional array and processed by the regressor network, which uses fully-connected layers to generate the output values. Only the extracted latent representation of the synthetic data is passed through this regression branch during training. If the experimental data were to pass through the regressor, then the concept of solving the task via simulated data only would be negated. The aim of the regressor is to try and predict the position of the spectrometer for each given sample, and the loss is calculated using the mean squared error. The network is then trained until the validation loss stagnates. The second phase of the training is the application of adversarial domain adaptation. The aim of this procedure is the alignment of the domains of the synthetic and experiment data so that the neural network performs well on experiment data even though it did not see this data during supervised training. For this task, both the latent representations of the synthetic and experiment data are fed into the discrimination branch, the discriminator tries to differentiate them and the inverse loss is fed back through the network via the gradient reversal layer. This inverse loss guides the training of the feature extractor in attempting to beat the discriminator and ultimately generate a domain-invariant representation of the data. The structure of the network and the flow of the different data streams is visualised in Figure \ref{network}.

\section{Results and Discussion}
Our results compare three stages of development and test performance on experimental data. The dataset comprises 1,620 image-label pairs acquired using the spectrometer at the UE52-SGM Beamline at BESSY II \cite{ue52}, with the spectrometer systematically scanned across the x-y plane at three primary z-axis positions, along with a smaller set of measurements at other z-values. The first test performs inference on the inverse surrogate model where no augmentation or domain adaptation has taken place, meaning the network was trained on pure simulation data. The second test is using a network trained with augmented simulated data, and the third introduces the use of domain adaptation.

\begin{figure}[H]
    \centering
    \includegraphics[width=0.7\linewidth]{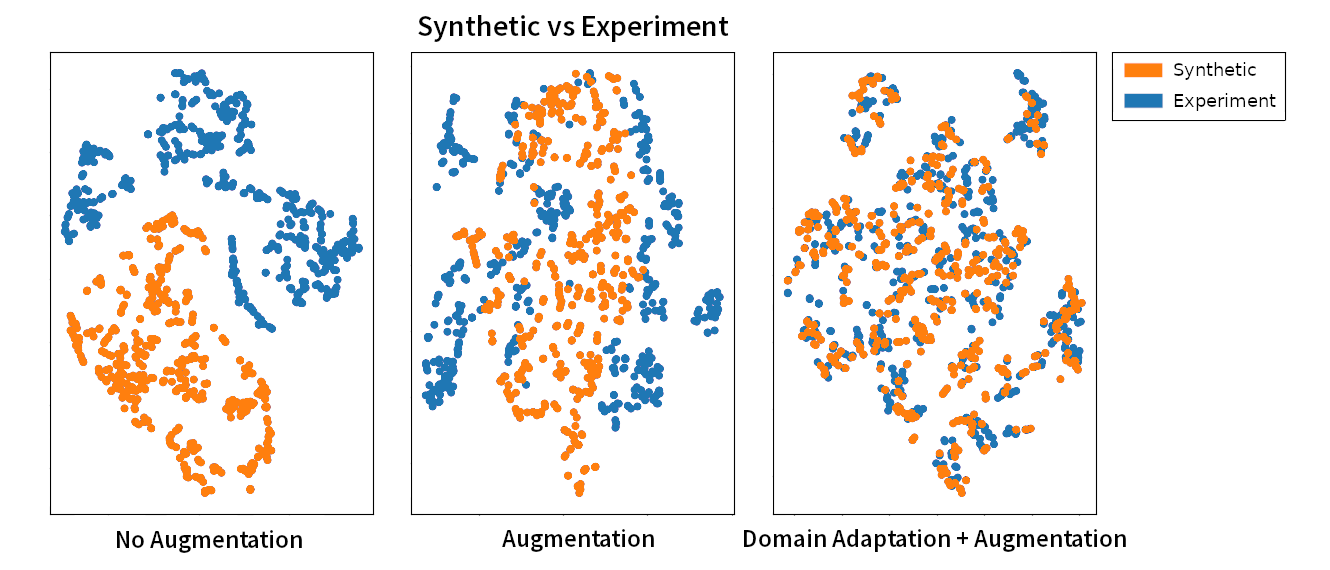}
    \caption{Visualisations of the latent representations whilst preforming inference using experiment data.}
\label{tsne}
\end{figure}

Analysis of the domain adaptation method has been undertaken using t-Distributed Stochastic Neighbor Embedding (t-SNE). t-SNE is a widely used method for visualising high-dimensional data in a lower-dimensional space. This method is particularly relevant for domain adaptation because it enables the visual assessment of how well features extracted from different domains align in the latent space. When applied to domain adaptation tasks, t-SNE helps to verify if the feature extractor has successfully learned to produce domain-invariant features. The visualisation of the simulation and experimental domains show no alignment. Even when augmentations are applied to the simulated data, the domains remain largely distinct. However, analysis of the domain adaptation shows that it is correctly aligning the synthetic and experimental domains in the latent feature space. This can be seen in Figure \ref{tsne}.

\begin{figure}[H]
    \centering
    \includegraphics[width=0.75\linewidth]{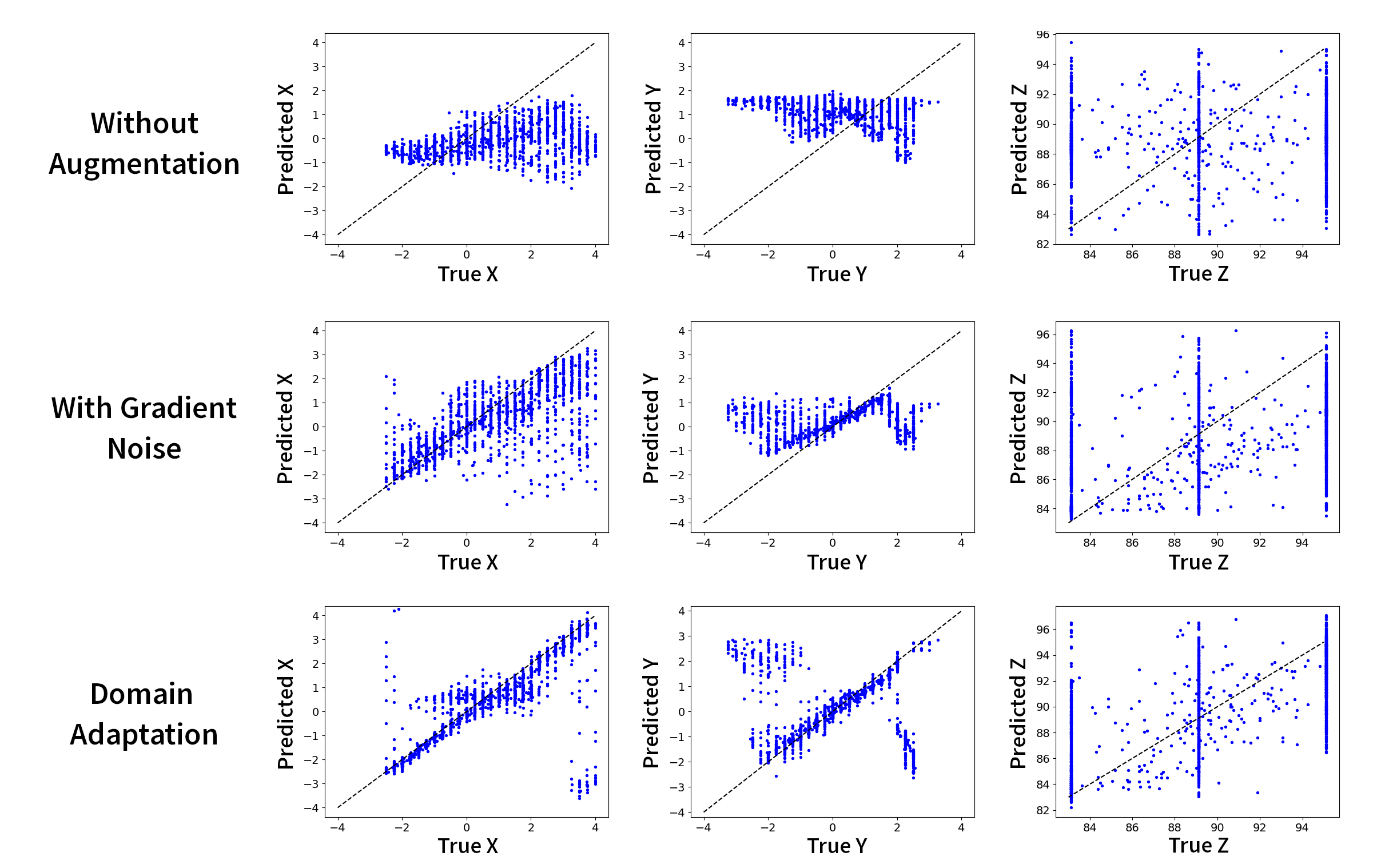}
    \caption{Inference results using 1,620 experimental images. True vs. predicted positions (x, y, and z) are shown, with rows representing no augmentation, data-specific augmentation, and augmentation with domain adaptation (top to bottom).}
\label{results}
\end{figure}

The goal is to predict the absolute positions of the RZP and detector relative to the sample in the spectrometer using three coordinates (\textit{x}, \textit{y}, \textit{z}). Figure \ref{results} compares the network's predictions with the true experimental values, where a diagonal would indicate perfect agreement between predicted and true values. The top row, showing the method without augmentation or adaptation, fails to correctly interpret all data points, predicting values close to the mean. The middle row, showing the simulation data with augmentations, is slightly improved. The network accurately predicts x- and y-axis positions for some data points, particularly those near the central position (0), likely due to stronger signal intensity. Incorporating augmented simulation data and domain adaptation (bottom row) significantly improves performance. This is most evident for the x-axis, where a clear diagonal trend can be seen. Even the y-axis demonstrates a noticeable improvement for data points in the central region, however data points around the -2 mm and +2 mm regions fail to improve and are in fact slightly further away. These inaccurate predictions primarily occur for images lacking axis-specific information, for example when the position of the slit obscures the signal. In these regions, the results appear worse because the network is no longer simply predicting mean values. However, the results have improved for useable data points which contain significant signal information which would contribute to aligning the spectrometer. Across all experiments, z-axis prediction is unsuccessful. As a zoom axis towards the camera, the z-axis has minimal impact on the signal's form, making its prediction challenging. Nonetheless, the domain adaptation results show definite improvement, providing validation for the approach.

\section{Conclusion and Outlook}
Our method builds on the adversarial domain adaptation techniques by Tzeng et al. \cite{8099799} and the gradient reversal layer by Ganin et al. \cite{ganin2016domainadversarialtrainingneuralnetworks}. It was improved upon by using augmentation techniques to close the domain gap and used in conjunction with a regressor network to predict target parameters. We have shown that domain adaptation, when combined with augmentation, performs significantly better than augmentation alone. In order to create a robust neural network trained with simulated data that can successfully operate on complex spectroscopy data, augmentations combined with domain adaptation show promising results. Our surrogate model inverts the simulation, mapping the image to input parameters, and successfully predicts the 
\textit{x}- and \textit{y}-axes, which correspond to the position, focus, and alignment of the fluorescence signal on the detector. The model can successfully operate on a small, unlabelled target dataset in order to infer parameters of our spectrometer, which can potentially see application when automating tasks such as the alignment of components during a beamtime.

The results obtained can still be improved upon, and we are currently working towards methods to achieve this. In particular, the training of models using higher resolution images, containing more information, and the creation of a dataset with a more diverse set of configurations is being undertaken. Success in this area would allow the scope of the work to be expanded beyond instrumentation to the task of aligning beamline components.

\section{Acknowledgments}
This research was funded in the framework of the Röntgen-Ångström Cluster (RÅC, https://www.rontgen-angstrom.eu/) on the German side via the Bundesministerium für Bildung und Forschung (BMBF, contract no. 05K20CBA) and on the Swedish side from the Swedish Research Council (grant agreement no. 2019-06093).

\printbibliography

\end{document}